\documentstyle[aps,pra,epsfig,twocolumn]{revtex}

\def\be{\begin{equation}}
\def\ee{\end{equation}}
\def\bea{\begin{eqnarray}}
\def\eea{\end{eqnarray}}
\def\bma{\begin{mathletters}}
\def\ema{\end{mathletters}}
\def\C{\hbox{$\mit I$\kern-.7em$\mit C$}}

\newcommand{\eins}{\mbox{$1 \hspace{-1.0mm}  {\bf l}$}}

\begin{document}
\draft

\title{Multiparticle entanglement and its experimental detection}

\author{W. D\"ur and J. I. Cirac}

\address{Institut f\"ur Theoretische Physik, Universit\"at Innsbruck,
A-6020 Innsbruck, Austria}

\date{\today}

\maketitle

\begin{abstract} 
We discuss several aspects of multiparticle mixed state entanglement and its 
experimental detection. First we consider entanglement between two particles 
which is robust against disposals of other particles. To completely detect these kinds of 
entanglement, full knowledge of the multiparticle density matrix (or of all 
reduced density matrixes) is required. Then we review the relation of the 
separability properties of $l$--partite splittings of a state $\rho$ to its 
multipartite entanglement properties. We show that it suffices to determine the 
diagonal matrix elements of $\rho$ in a certain basis in order to detect 
multiparticle entanglement properties of $\rho$. We apply these observations to 
analyze two recent experiments, where multiparticle entangled states of 3 (4) particles were 
produced. Finally, we focus on bound entangled states (non--separable, 
non--distillable states) and show that they can be activated by joint actions of 
the parties. We also provide several examples which show the activation of bound 
entanglement with bound entanglement. 
\end{abstract}


\narrowtext


\section{Introduction}

Entanglement is at the heart of Quantum Information theory. In recent years, 
there has been an ongoing effort to characterize quantitatively and 
qualitatively entanglement. While for bipartite systems essential parts of 
this problem are solved, many questions remain still open for multipartite 
systems. In this case, there exist several possible approaches to identify 
different kinds of multiparticle entanglement (MPE), and many interesting 
phenomena related to MPE have been discovered 
\cite{effects,Du99a,Sm00,Du00,Sh00,Du00M}. 

In this work, we review some possible approaches to identify different kinds of 
MPE and discuss its experimental detection.

\subsection{Bipartite Entanglement, Separability and Distillability}\label{sepdis}

Let us start with the simplest case of bipartite systems and review some basic 
concepts related to bipartite entanglement. Let $A$ and $B$ be two spatially 
separated systems of dimension $d_A$ [$d_B$] respectively. A state $\rho$ is 
said to be separable if it can be written as a convex combination of product 
states, i.e.
\be 
\rho = \sum_i p_i|a_i\rangle_{A}\langle a_i| \otimes |b_i\rangle_{B}\langle b_i|.\label{sepf}
\ee 
In case this is not possible, $\rho$ is said to be entangled. Note that 
separable states $\rho$  are states which can be prepared locally by the 
parties, i.e. $\rho$ is only classically correlated. As inseparable (entangled) 
states are very interesting, both from a fundamental and from a practical point 
of view, one of the main problems in Quantum Information Theory is the problem 
of establishing whether a given state $\rho$ is separable or not. We have that 
condition (\ref{sepf}) is in general very difficult to check, as there exist (in 
general) infinitely many ways to write a given density operator $\rho$ as a 
convex combination of (possible entangled) pure states. However, the problem of 
separability has been extensively studied in recent years \cite{Le00}, and in 
the case of two qubits ($d_A=d_B=2$), necessary and sufficient conditions for 
separability have been obtained \cite{Pe96,Ho96}.  
In particular, for two qubits one can use the partial transposition criterium 
\cite{Pe96,Ho96} which states that (i) $\rho$ is separable iff $\rho^{T_A} \geq 0$ 
\cite{Ho96}.  Here, $T_A$ denotes transposition in $A$ in a given orthonormal 
basis $S_A=\{|k\rangle\}_{k=1}^{d_A}$, and $X\geq 0$ means that all eigenvalues 
of $X$ are $\geq 0$. For higher dimensional systems ($d_A,d_B > 2$), positivity 
of the partial transposition is only a necessary, but not sufficient condition 
for separability.

For inseparable (entangled) density operators $\rho$, one may also ask whether 
the entanglement contained in $\rho$ can be distilled. That is, whether out of 
(arbitrary) many copies of $\rho$, a maximally entangled state (MES) such as the 
singlet state $|\Psi^-\rangle=(|01\rangle-|10\rangle)/\sqrt{2}$ shared by the 
parties $A$ and $B$ can be created by means of local operations and classical 
communication. In case this is possible, $\rho$ is said to be distillable. 
Again, for two qubits it turns out that the partial transposition provides a 
necessary and sufficient condition for distillability: (ii) $\rho$ is 
distillable iff $\rho^{T_A} \not\geq 0$ \cite{Ho97}. For higher dimensional 
systems ($d_A,d_B > 2$), non positive partial transposition is a necessary, but 
not sufficient condition for distillability.

The partial transposition criterium (i) and (ii) thus provides a necessary and 
sufficient condition for separability and distillability for two qubit systems.

\subsection{Multiparticle entanglement}

The aim of this paper is to extend these ideas to multiparticle systems, in 
particular to study separability and distillability properties of multiparticle 
systems. However, there are various aspects of multiparticle entanglement. For 
example, there exist obviously many different kinds of entanglement in a 
multiparticle system, as one may have bipartite entanglement shared by, say, 
parties $A_1$ and $A_2$ as well as bipartite entanglement shared by two other 
parties, say $A_2$ and $A_3$. In addition, there exist true $N$--partite 
entanglement, for example MES of $N$ particles such as the 
Greenberger-Horne-Zeilinger (GHZ) state \cite{Gr89} 
\be
|GHZ\rangle=\frac{1}{\sqrt{2}}(|0^{\otimes N}\rangle+|1^{\otimes N}\rangle) \label{GHZ}.
\ee

Concerning for example the question of distillability, one may 
consider distillability of bipartite entanglement between pairs of particles or 
of true $N$--partite entanglement between a group of particles. In both cases, 
one may either ignore the remaining particles or allow them to assist the other 
parties in order to distill a MES. On may also consider partitions of the 
system, i.e. allowing some of the parties to act together and perform joint 
operations, and determine the distillability (and separability) properties with 
respect to this partitions, which in turn provide information about the 
entanglement properties of the whole system. Each of the situations just 
described is concerned with a different aspect of multiparticle entanglement, 
and will be discussed in more detail in the following.

From an experimental point of view, it is of particular interest to detect 
whether a $N$--particle state is distillable to a MES of $N$--particles. We will 
provide a simple, sufficient criteria which allows ---without full knowledge of 
the density matrix--- to detect true $N$--qubit entanglement. In addition, this 
criteria allows to detect different kinds of multiparticle entanglement as well. 
We also observe that there exist more kinds of multipartite entanglement then 
the obvious ones already mentioned previously (all possible combinations of 
maximally entangled $l$--partite states for different $l$). In particular, we 
consider bound entangled states, i.e. non separable, non distillable states and 
show that they can be activated under certain circumstances. We provide examples 
illustrating quite surprising effects related to bound entanglement and its 
activation.

This paper is organized as follows. We start in Sec. \ref{Molecules} by 
discussing bipartite aspects of MPE, that is entanglement which is robust 
against disposal of particles. We discuss the necessary information which is 
required to detect these aspects of MPE. In Sec. \ref{splittings}, we choose a 
different approach and concentrate on $l$--partite aspects of MPE. Using 
$l$--partite splittings of the system, we show how to completely determine the 
separability and distillability properties of a certain family of states, i.e 
its MPE properties. Using these results, we provide a simple (sufficient) 
criteria to (experimentally) detect different kinds of MPE. We illustrate this 
method by applying it to two recent experiments, where MES of 3 (4) particles 
respectively were created. Finally, in Sec. \ref{bound} we focus on an 
interesting phenomena related to MPE, namely on bound entanglement and its 
activation. In particular, we show that bound entanglement can sometimes be 
activated by joint actions of some of the parties or alternatively with help of 
a different kind of bound entanglement. We give several examples to illustrate 
these effects.


\section{Entanglement which is robust against disposal of particles}\label{Molecules}

In this Section, we concentrate on bipartite aspects of multipartite entanglement, in 
particular on bipartite entanglement which is robust against disposal of 
particles. We consider $N$ spatially separated parties $A_1,\ldots,A_N$, each 
possessing a qubit. 

We say that two particles are (bipartite) entangled if their reduced density 
operator\footnote{Given a $N$--partite state $\rho$, the reduced density 
operator $\rho_{12}$ of party $A_1$ and $A_2$ is defined as $\rho_{12}\equiv 
{\rm tr}_{3,\ldots,N} (\rho)$. The operator $\rho_{12}$ is separable if it can 
be written as a convex combination of product states.} is non--separable, i.e. 
the two particles share entanglement, independent what happens to the remaining 
particles. When considering the 
reduced density operator of two parties, we deal with the situation where the 
information about all remaining particles is not accessible (or the remaining 
parties are not willing to cooperate). 
Such a definition is very suitable from a practical point of view, as there are 
certain multipartite scenarios where one is interested in entanglement 
properties of pairs of parties, which are independent of other parties. In 
addition, in certain experiments one may be faced with such a situation, e.g. 
when one of the particles escapes from a trap. The remaining particles should then be 
described by the reduced density operator.
Note that in this sense, the GHZ state (\ref{GHZ}) contains no (bipartite) 
entanglement at all, as all reduced density operators are separable. However, 
the GHZ state can be regarded as MES of $N$ particles in several other senses 
\cite{Gi98}.


\subsection{Entanglement molecules}

In \cite{Du00M}, it was shown that there exist $N$--particle states which are 
still entangled when tracing out {\it any} $(N-2)$ particles, i.e. there states 
where all particles are entangled with all other particles. In addition, it was 
shown there that there exist $N$--partite states $\rho$ where one can choose for 
each of the $N(N-1)/2$ reduced density operators $\rho_{kl}$ {\it independently} 
whether it should be separable or inseparable.  This allows to build general 
structures of $N$ particle states, which were called 'Entanglement molecules' in 
\cite{Du00M}.

The following family of $N$ qubit states includes all possible 
configurations of 'Entanglement molecules' \cite{Du00M}. First we specify for each of the 
reduced density operators $\rho_{kl}$ whether it should be distillable or not \cite{noteMol}, 
i.e. whether entanglement between the parties $A_k$ and $A_l$ can be distilled 
--- without help of the remaining parties --- or not. Let $I=\{k_1l_1, \ldots 
,k_Ml_M\}$ be the set of all those pairs where distillation should be possible, 
i.e. for $kl\in I$, we have that $\rho_{kl}$ is distillable. We define the 
state 
\be 
|\Psi_{ij}\rangle \equiv |\Psi^+\rangle_{ij} \otimes |0\ldots 0\rangle_{\rm rest}, 
\ee 
that is the particles $A_i$ and $A_j$ are in a MES, namely 
$|\Psi^+\rangle=1/\sqrt{2}(|01\rangle+|10\rangle)$, and the remaining particles 
are disentangled from each other and from $A_iA_j$. The family of 
states
\be
\rho_I= \frac{1}{M} \sum_{kl \in I}  x_{kl}|\Psi_{kl}\rangle\langle\Psi_{kl}|,\label{familyM}
\ee
has the desired properties, which can be checked \cite{Du00M} by calculating the 
reduced density operators $\rho_{kl}$ and using the partial transposition 
criterium. We have that $M \equiv \sum_{kl \in I} x_{kl}$ is a normalization 
factor. The bipartite aspects of multipartite entanglement were also analyzed 
in \cite{Wo00}.


\subsection{Experimental detection}

Given a $N$--qubit state $\rho$, how can we determine its (bipartite) 
entanglement properties ? One possibility is to completely determine the 
$N$--partite density matrix of $\rho$. Given $\rho$, one can easily calculate 
all possible reduced density operators $\rho_{kl}$ and determine the 
separability properties of each $\rho_{kl}$. Due to the fact that we deal with 
qubits, one can use the partial transposition criterium (see Sec. \ref{sepdis}) 
to determine for each of the reduced density operators $\rho_{kl}$ whether it is 
separable or distillable. In case $\rho_{kl}$ is inseparable, a MES shared by 
the parties $A_k$ and $A_l$ can be distilled.

However, it is rather difficult to completely determine the density matrix of 
$N$--qubit system, which is required in the procedure described above. 
Alternatively, one can concentrate from the very beginning on the properties of 
the reduced density operators $\rho_{kl}$, i.e. ignoring the remaining particles 
and just measuring the bipartite density operator $\rho_{kl}$. In this case, all 
$N(N-1)/2$ different reduced density operators have to be determined 
independently and can then be analyzed using the partial transposition 
criterium. 

Still, it might be too demanding to completely determine the density matrix of a 
two qubit system, which is necessary to completely determine the separability 
properties of this system. However, in order to {\it detect} entanglement in a 
two--qubit system, it suffices to show that the Fidelity $F$, i.e. the overlap 
with an arbitrary MES, fulfills $F>1/2$. Note however that this is a sufficient 
condition for inseparability (distillability), which is in general not 
necessary. So one can alternatively measure the overlap of each of the reduced 
density operators $\rho_{kl}$ with a MES. Observing that for a given $\rho_{kl}, 
F>1/2$ implies that out of $\rho_{kl}$ a MES shared among $A_k$ and $A_l$ can be 
distilled. However, when one finds $F\leq 1/2$, nothing can be concluded about 
the separability properties of $\rho_{kl}$.

Establishing the (bipartite) entanglement properties of a state $\rho$ is 
however not the only possibility to determine the multipartite entanglement (MPE)
properties of $\rho$. The bipartite entanglement properties, i.e. the properties 
of the reduced density operators $\rho_{kl}$, are only a certain aspect of the 
MPE properties of $\rho$. There are other aspects of 
MPE and alternative ways to detect the presence of 
MPE, which will be discussed in the next Section. 


\section{$l$--partite aspects of multiparticle entanglement}\label{splittings}

In this section we first review the concepts of $l$--partite splittings, 
$l$--separability and distillability. These properties can be used to completely 
characterize the multiparticle entanglement properties of an arbitrary mixed 
state $\rho$ \cite{Du99c}. We then review the properties of a family of 
$N$--qubit states $\rho_N$ introduced in Ref. \cite{Du99a} and completely 
determine the entanglement properties of this family. Finally we show that these 
results can be used to determine entanglement properties of general states 
$\rho$ without complete knowledge of the density matrix. In particular, it 
suffices to determine the diagonal matrix elements of $\rho$ in a certain basis 
in order to establish sufficient conditions for the presence of multipartite 
entanglement.  We provide a simple receptive to experimentally detect different 
kinds of multipartite entanglement. We apply the results to two recently 
performed experiments \cite{Sa00,Ra00} to illustrate the usefulness of our 
method.

\subsection{Bipartite and $l$--partite splittings}\label{bip}

Let us denote by ${\cal P}$ the set of all possible bipartite splittings of 
$N$ parties into two groups. For example, for $3$ parties ${\cal P}$ contains 
the splittings $(A_1A_3)$--$(A_2)$, $(A_2A_3)$--$(A_1)$, and 
$(A_3)$--$(A_1A_2)$. We will denote these bipartite splittings by $P_{k}$, where 
$k=k_1k_2\ldots k_{N-1}$ is a chain of $N-1$ bits, such that $k_n=0,1$ if the 
$n$--th party belongs to the same group as the last party or not. For example, 
for $3$ parties the splittings $(A_1A_3)$--$(A_2)$, $(A_2A_3)$--$(A_1)$, and 
$(A_3)$--$(A_1A_2)$ will be denoted by $P_{01}$, $P_{10}$, and $P_{11}$, 
respectively. We will denote by $A$ the side of the splitting to which the party 
$N$ belongs and by $B$ the other side. In a similar way, one can consider $l$--partite splittings 
$S_l$, where the parties form exactly $l$ groups. In the following, when we 
consider $l$--partite splittings, the parties in each of the $l$ groups will be allowed 
to act together (i.e. to perform joint operations).

\subsection{$l$--separability and distillability}

Here we review the notion of separability and distillability in the case of
multiparticle systems. We consider $N$ parties, each holding a system with
dimension $d_i$, i.e. ${\cal H}=\C^{d_1}\otimes \ldots \otimes\C^{d_N}$. We call
$\rho$ fully separable if it can be written as a convex combination
of (unnormalized) product states, i.e. 
\be \rho = \sum_i |a_i\rangle_{\rm
party1}\langle a_i| \otimes |b_i\rangle_{\rm party2}\langle b_i|\otimes \ldots
\otimes |n_i\rangle_{{\rm party} N}\langle n_i|.\label{sepa1} 
\ee 
In the following, we will consider a system of $N$ qubits, each hold by one of 
the parties $A_1,A_2,\ldots,A_N$. In this case, $d_1=d_2=\ldots d_N=2$. A state 
$\rho$ is called $k$--separable with respect to a specific $k$--partite 
splitting iff it is fully separable in the sense that we consider $\rho$ as a 
$k$--party system, i.e. as a state in ${\cal H}=\C^{d_1}\otimes \ldots 
\otimes\C^{d_k}$. In order to completely determine the separability properties 
of a state $\rho$, one should determine the separability properties of {\it all} 
possible $l$--partite splittings for all $l\leq N/2$. Based on this information, 
one can establish an hierarchic classification of the entanglement properties of 
$\rho$ (see Ref. \cite{Du99c} for details). It turns out that the separability 
properties of the different $l$--partite splittings for different $l$ are not independent of each other, 
which strongly simplifies the classification and reduces the number of possible 
classes. In some cases we will deal with in the following, it even suffices to 
determine the biseparability properties of a state, i.e. to establish the 
separability properties of all possible bipartite splittings. This is due to the 
fact that the $l$--separability properties in this case are completely 
determined by the biseparability properties of $\rho$. 

In a similar way, one can establish the distillability properties of a state 
$\rho$. Given a bipartite splitting $P_k$, a state $\rho$ is called distillable 
with respect to the splitting $P_k$, if ---out of $N$ identical copies of 
$\rho$--- the two groups $A$ and $B$ (which correspond to the two groups of the 
splitting) can create by means of local operations and classical communication a 
MES such as $|\Phi^+\rangle=1/\sqrt{2}(|00\rangle+|11\rangle)$, shared among $A$ 
and $B$. Recall that the term ''local'' in this case refers to local operation 
with respect to the  groups $A$ and $B$, but may involve joint operation on the 
particles within one group. In the case of distillability, it is not necessary 
to consider $l$--partite splitting and the possible creation of $l$--party GHZ 
states, as the creation of pairwise entanglement between any two out of $l$ 
parties is a necessary and sufficient condition for the distillation of a 
$l$--partite GHZ state shared among those parties \cite{Du99c}. However, one may 
ask whether two subgroups ---not containing all parties--- are capable of 
distilling a MES with help of the remaining parties. For a certain family of 
states, we will give necessary and sufficient conditions when this is possible.

\subsection{Family of states $\rho_N$}\label{family}

Let us consider $\rho_N$, the family of $N$--qubit states introduced in
\cite{Du99a}. We have that $\rho\in\rho_N$ if it can be written as
\bea
\label{rhoN}
\rho &=& \sum_{\sigma=\pm} \lambda_0^\sigma |\Psi^\sigma_0\rangle\langle
  \Psi^\sigma_0| \nonumber\\
&& + \sum_{k\not= 0} \lambda_k (|\Psi^+_k\rangle\langle \Psi^+_k|
  + |\Psi^-_k\rangle\langle \Psi^-_k|),
\eea
where
\be\label{GHZbas}
|\Psi^\pm_k\rangle \equiv \frac{1}{\sqrt{2}} (|k_1k_2\ldots k_{N-1} 0\rangle \pm
|{\bar k_1}{\bar k_2}\ldots {\bar k_{N-1}}1\rangle),
\ee
are GHZ--like states with $k=k_1k_2\ldots k_{N-1}$ being a chain of $N-1$ bits,
and ${\bar k_i}=0,1$ if $k_i=1,0$, respectively. We have that $\rho_N$ is
parameterized by $2^{N-1}$ independent real numbers. The labeling is chosen such
that $\Delta\equiv \lambda_0^+- \lambda^-_0 \ge 0$. As we will see below, both
the separability and distillability properties of the states belonging to this
family are completely determined by the coefficients
\be
\label{sk}
s_{k}\equiv \left\{\begin{array}{l}
\mbox{1 if $\lambda_k < \Delta/2$} \\
\mbox{0 if $\lambda_k\geq \Delta/2$.}
\end{array}
\right.
\ee
Let us emphasize that the notation used for the states of this family parallels 
the one used to denote the partitions $P_k$, i.e. there is a one to one 
correspondence between $P_k$ and $s_k$. Note that there are no restrictions to 
the values of these coefficients; that is, for any choice of $\{s_k\}$ there 
always exists a state $\rho\in\rho_N$ with these values.
We will now summarize the properties of 
states belonging to the family (\ref{rhoN}) \cite{Du99c,Du00}: 

{\bf (i) Depolarization:} An arbitrary state $\rho$ can be depolarized to the standard form 
(\ref{rhoN}) by a sequence of $N$--local operations while keeping the values of 
$\lambda_0^\pm\equiv \langle \Psi^\pm_0|\rho|\Psi^\pm_0\rangle$ and $2\lambda_j 
\equiv \langle \Psi^+_j|\rho|\Psi^+_j\rangle + \langle 
\Psi^-_j|\rho|\Psi^-_j\rangle$ unchanged \cite{Du99c}. 

{\bf (ii) Separability:} For any bipartite splitting $P_k \in {\cal P}$, and $\rho \in \rho_N$ 
we have $\rho^{T_{A}} \geq 0 \Leftrightarrow s_k=0 \Leftrightarrow \rho$ is 
separable with respect to this splitting\footnote{$\rho^{T_{A}}$ denotes the 
partial transposition with respect to the parties $A$. For the definition of 
partial transposition in multiparticle systems see \cite{Pe96,Du99c}. The 
relation between subsystem $A$ and $P_k$ is given in Sec. \ref{bip}.} 
\cite{Du99c}. More generally, $\rho\in\rho_N$ is $l$--separable with respect to 
a specific $l$--partite splitting $S_l$ iff all bipartite splittings $P_k$ which 
contain\footnote{A $l$--partite splitting $S_l$ is contained in a $k$--partite splitting 
$P_k$ iff $P_k$ can be obtained from $S_l$ by joining some of the parties of 
$S_l$.} $S_l$ are separable (have $s_k=0)$.

{\bf (iii) Distillability:} Let $\rho \in \rho_N$, $C=\{A_{i_1},\ldots,A_{i_M}\}$ and 
$D=\{A_{j_1},\ldots,A_{j_L}\}$ be two disjoint groups of $M$ and $L$ parties 
respectively, whereas the rest of the parties are separated. A MES between $C$ 
and $D$ can be distilled iff $\rho$ is non--separable with respect to all those 
bipartite splittings $P_k$ in which the groups $C$ and $D$ are located on 
different sides (i.e. all corresponding $s_k=1$). It follows that $\rho$ is 
distillable with respect to a bipartite splitting $P_k$ $\Leftrightarrow s_k=1$ 
\cite{Du00}.

Note that (ii) and (iii) {\it completely} determine the separability and 
distillability properties of an arbitrary state $\rho\in\rho_N$ and thus the 
multipartite entanglement properties of this state. We also have that (iii) 
already implies complete knowledge about the distillability of $k$--partite GHZ 
states, as the creation of pairwise entanglement between any two out of $k$ 
parties is a necessary and sufficient condition for the distillation of a 
$k$--partite GHZ state shared among those parties \cite{Du99c}.

\subsection{Implications for experimental detection of multipartite entanglement}

We have that (i-iii) together provide a simple criterium for the detection of 
multipartite entanglement for arbitrary mixed states $\rho$: From (i) follows 
that any state $\rho$ is {\it at least} as entangled as the depolarized version 
$\tilde\rho\in\rho_N$ of $\rho$. This is due to the fact that a sequence of 
local operations may destroy some entanglement, but cannot create any new kind 
of entanglement which was not present in the initial state. This already gives 
us a receptive to detect different kinds of multipartite entanglement of an 
arbitrary state $\rho$:
\begin{itemize}
\item Determine the following diagonal matrix elements of $\rho$: 
\bea 
\lambda_0^\pm &\equiv& \langle \Psi^\pm_0|\rho|\Psi^\pm_0\rangle \nonumber \\ 
2\lambda_j &\equiv& \langle \Psi^+_j|\rho|\Psi^+_j\rangle + \langle 
\Psi^-_j|\rho|\Psi^-_j\rangle  \\
&=& \langle j0|\rho|j0\rangle +\langle \bar j1|\rho|\bar j1\rangle \nonumber.
\eea
Note that determining $\lambda_0^\pm$ requires a measurement in an entangled 
basis (GHZ basis), while determining $\lambda_j$ corresponds to a measurement in 
a product basis. Recall that $|j0\rangle =|j_1j_2\ldots j_{N-1} 0\rangle$ and 
$|\bar j0\rangle =|\bar j_1 \bar j_2\ldots \bar j_{N-1} 1\rangle$ (see 
(\ref{GHZbas})). Equivalently, it suffices to determine all {\it diagonal} 
matrix elements of $\rho$ in the standard basis plus one off--diagonal element, 
namely $|0\ldots 0\rangle\langle1\ldots 1|$.  

\item Calculate $\Delta= \lambda_0^+-\lambda_0^-=2 Re(\langle 0\ldots 
0|\rho|1\ldots 1\rangle)$ and determine the coefficients $s_k$ given in 
(\ref{sk}). If at least one $s_k=1$, we have that $\rho$ is entangled.

\item Use (i-iii) to determine the (minimal) entanglement properties of the 
state $\rho$. Note that obtaining $s_k=0$ for a certain bipartite splitting 
$P_k$ does not imply that $\rho$ is separable with respect to this splitting. It 
might well be that $\rho$ is inseparable (entangled) with respect to $P_k$, but 
the corresponding depolarized state $\tilde\rho$ is separable. However, 
obtaining $s_k=1$ ensures that a certain kind of entanglement is present in the 
state $\rho$ -- namely that $\rho$ is inseparable with respect to the bipartite 
splitting $P_k$. In particular, one can distill a GHZ state from $\rho$ iff 
$s_k=1 \forall k$. 

\end{itemize}

\subsection{Application to recent experiments}

Let us apply this method to two recent experiment performed by Sacket et al. 
\cite{Sa00} and Rauschenbeutel et al.\cite{Ra00}. 

In \cite{Sa00}, the creation of an (mixed) entangled state $\rho$ of 4 ions, 
whose overlap with the GHZ state $|\Psi_0^+\rangle$ is $F=0.57 \pm 0.02$, was 
reported. It was argued that it is sufficient to obtain $F>1/2$ in order to be 
sure that the state is 4--partite entangled. This sufficient criteria is however 
---in some cases--- much too demanding and can be relaxed using the results 
presented in this work. Imagine for example that the 4 qubit state $\rho$ is of 
the form 
\be
\rho(x)=x|\Psi_0^+\rangle\langle\Psi_0^+|+\frac{1-x}{16}\eins_4
\ee
This is clearly a special case of the state $\rho_4$ with 
$\lambda_0^-=\lambda_j=\frac{1-x}{16}$, $\lambda_0^+=x+\frac{1-x}{16}$ and thus 
$\Delta=x$. Using (ii) and (iii), we can state that $\rho(x)$ is fully 
non--separable and distillable to a 4 party GHZ state state iff $x>1/9$, which 
corresponds to $F>1/6$ \cite{Du99c}. Note that the bound $F>1/2$ ---which is 
independent of the number of parties $N$--- corresponds to a worst case 
scenario, where it is assumed that $\lambda_0^+=F$ and the remaining weight is 
distributed on $\lambda_0^-$ and one specific $\lambda_k$. In this case, we have 
for $F>1/2$ that $\Delta > 2\lambda_k \forall k$. If the remaining weight $(1-F)$ 
is however distributed on $\lambda_0^-$ and more than one $\lambda_k$, it 
automatically follows that $\Delta > 2\lambda_k \forall k$ is already fulfilled 
for all $\lambda_0^+\equiv F>F_0$, where $F_0<1/2$. The weakest bound on the 
Fidelity $F$ can be obtained by assuming that the state $\tilde\rho_N$ is of the 
following form: $\lambda_0^+=F, \lambda_0^-=0$ and $2\lambda_k=(1-F)/(2^N-2)$. 
This ensures that $\tilde\rho_N$ has $\Delta > 2\lambda_k \forall k$ and is thus 
distillable to a $N$--party GHZ state iff $F>1/(2^N-1)$. For $N=4$, we obtain 
$F>1/15$. We thus have that additional knowledge of the shape of the state may 
relax the necessary conditions to ensure that a state is entangled.  

Let us now focus on the specific experiment \cite{Sa00} and apply these 
observations.  Unfortunately, the published experimental data is not sufficient 
to determine all coefficients $\lambda_k$. However, one can easily determine 
\bea
\lambda_0^\pm&=&1/2(\langle0000|\rho|0000\rangle + 
\langle1111|\rho|1111\rangle)\nonumber \\&&\pm Re(\langle 0000|\rho|1111\rangle)  =  0.35\pm 
0.215 (\pm 0.02),
\eea
from which follows that $\Delta=0.43 (\pm 0.02)$. In addition, one can also 
bound the other coefficients $\lambda_k$ and finds \cite{noteExp}  
\bea
0 \leq 2\lambda_k \leq 0.2 (\pm 0.04) &\hspace{5pt}{\rm iff }\hspace{5pt}& k\in\{001,010,100,111\} \nonumber\\
0 \leq 2\lambda_k \leq 0.1 (\pm 0.02) &\hspace{5pt}{\rm iff }\hspace{5pt}& k\in\{011,101,110\}.
\eea
We thus have that $\Delta > 2\lambda_k \forall k$ as expected. Note however that 
a Fidelity $F<1/2$ would have been sufficient to ensure that the produced state is 
truly 4--partite entangled. Assume for example that white noise is added to the 
experimentally produced state, i.e. $\tilde\rho=x\rho+(1-x)/16\eins_4$. Using the 
bounds on $\lambda_k$ just derived, we find that $\tilde\rho$ remains truly 
4--partite entangled for $x>0.58559$, which corresponds to a Fidelity of 
$F>0.3597$, significantly below 1/2. 

Our method should thus simplify the task to detect an entangled state of a 
larger number of particles ($N>4$), as it relaxes the necessary conditions for 
the detection of true $N$--partite entanglement. Note that it would be highly 
desirable to measure all diagonal coefficients in the standard basis 
independently rather than Projections into subspaces $P_j$ with $j$ particles in 
$|0\rangle$ and $N-j$ particles in $|1\rangle$ as done in \cite{Sa00}. Doing so, one could determine 
the coefficients $\lambda_k$ directly and does not have to use a ''worst case 
scenario'' in order to establish bounds on $\lambda_k$ as we did here (see 
\cite{noteExp}). In addition, different kinds of entanglement which do not 
correspond to $N$--party GHZ entanglement can be detected as well. In the next 
section, we show that states showing these different kinds of entanglement may 
also be interesting to produce, as they provide examples for surprising 
effects such as the activation of bound entanglement.  

One may also adopt this method to other experiments, such as the one performed 
by Rauschenbeutel et al. \cite{Ra00}, where a maximally entangled state of three 
spin $\frac{1}{2}$ systems (two atoms plus one cavity mode) was created. Let us 
first adopt the notation used in \cite{Ra00} to the one used throughout this 
paper: $|+_j\rangle = |1\rangle,  |-_j\rangle = -|0\rangle$, where e.g. 
$|+_1\rangle=|e_1\rangle, |+_2\rangle=(|g_2\rangle+|i_2\rangle)/\sqrt{2}$ and 
$|+_C\rangle=|g_3\rangle$ (see Equ. (3) and below in \cite{Ra00}). It follows 
that the longitudinal correlations given in Fig. 3 of \cite{Ra00} correspond 
to the diagonal matrix elements of $\rho$ in the basis (from left to right) 
$\{|011\rangle,|010\rangle,|001\rangle,|000\rangle,|111\rangle,|110\rangle,|101\rangle
,|100\rangle\}$. From this we can determine
\bea
2\lambda_{01}&=&0.14 (\pm 0.04) \nonumber \\
2\lambda_{10}&=&0.155 (\pm 0.04)  \\
2\lambda_{01}&=&0.128 (\pm 0.04) \nonumber 
\eea
From the transverse correlations we find 
\be
\Delta=2 Re (\langle 000|\rho|111\rangle=2 V_\bot =0.28 (\pm 0.04).
\ee
Thus we have that $\Delta > 2\lambda_k \forall k$ and we can conclude that the 
experimentally detected state $\rho$ is in fact distillable to a 3--party GHZ 
state. Note that in \cite{Ra00}, it was necessary to take known detection errors 
into account in order to obtain $F>0.5$. Here we can state that even without 
taking these errors into account, the state $\rho$ is true tripartite entangled, 
although its fidelity $F=0.43 <1/2$.


\section{Bound entanglement and its activation}\label{bound}

Let us now consider $N$ spatially separated parties, $A_1,\ldots,A_N$, who share 
$M$ identical copies of a $N$--qubit state $\rho$, where $M$ can be as large as 
we wish. This ensures that the parties can use distillation protocols 
\cite{Be96} in order to obtain MES between some of them. In case this is 
possible, we say that the state $\rho$ is distillable (with respect to the 
specific parties which obtain the MES). If no MES shared between any two of the 
parties can be distilled and in addition the state $\rho$ is not fully separable 
(i.e. entangled), we say that $\rho$ is bound entangled (BE).


\subsection{Activating bound entanglement by joint actions}

Given a bound entangled state (BES), in some cases it is possible to activate 
the bound entanglement. We say that a BES can be
activated if it becomes distillable once some of the parties join and form
groups to act together. Note that instead of allowing some parties to join we
could have allowed them to share some extra MES. In that case we would have the same
situation given the fact that separated parties sharing MES can perform any
arbitrary joint operation by simply teleporting\cite{Be93} back and forth the states of
their particles.

The first example of this kind was given in 
\cite{Du99a}. There it was shown that given a certain BES shared by 3 parties, 
providing some extra bipartite entanglement between $A_1$ and $A_2$ enables the 
3 parties to create a tripartite $GHZ$ state. 

In \cite{Sm00}, Smolin presented another example of this kind involving 4 parties. 
This example has the additional feature that only a single copy of a BES $\rho$ 
is required in order to distill a MES shared by two of the parties (say $A_1$ 
and $A_2$) once the other two parties (say $A_3$ and $A_4$) are allowed to act 
together and perform joint operations. 

Using states of the form (\ref{rhoN}), several examples showing the activation 
of different kinds of BE by joint actions of some of the parties were provided 
in \cite{Du00}. In addition, a systematic way for the construction of different 
kinds of activable BES was provided there. Let us review some of the examples 
given in \cite{Du00}:

\noindent
{\bf Example I:} The state $\rho_I$ becomes distillable iff the parties form two
groups with exactly $j$ and $N-j$ members, respectively. Furthermore, it does
not matter which of the parties join in each group, but only the number of
members. For example, if $N=8$ and $j=3$, we have that
$\rho_I$ is distillable if exactly $3$ and $5$ parties join, but remains
undistillable when the parties form two groups with 1-7, 2-6, 4-4 members, or if
they form more than two groups. In particular, $\rho_{I}$ is not distillable if
the parties remain separated from each other, which corresponds to having $8$
groups.
We can take as state $\rho_{I}$ one from the family $\rho_N$ which has $s_{k}=1$ iff the 
number of ones in $k$ is $j$ or $(N-j)$ and $s_{k}=0$ otherwise (this 
means that all bipartite splittings which contain exactly $j$ members in one 
group are distillable, and all others are separable).

\noindent
{\bf Example II:} The state $\rho_{II}$ becomes distillable iff the parties form 
two groups, where the first group contains a {\it specific} set of $L$ parties 
$A=\{A_{k_1},\ldots A_{k_L}\}$, and the second group contains the remaining 
parties. For all other configurations in groups $\rho_{II}$ remains 
undistillable. For example, we have for $N=5$ and $A=\{A_1,A_3,A_5\}$ that 
$\rho_{II}$ is distillable iff the the parties form two groups, 
$(A_1A_3A_5)-(A_2A_4)$, and not distillable otherwise. We can take $\rho_{II} 
\in \rho_N$ such that $s_k=1$ only for one specific $P_k$. For $N=5$, 
choosing $s_{0101}=1$ ensures that $\rho_{II}$ is inseparable and thus 
distillable with respect to the bipartite 
splitting $(A_1A_3A_5)-(A_2A_4)$ and separable (and thus undistillable) 
otherwise.

\noindent
{\bf Example III:} $\rho_{III}$ is a BES of $N=4$ parties for which once the 
parties $(A_3A_4)$ form a group, a GHZ--like state can be distilled among $A_1$, 
$A_2$, and the group $(A_3A_4)$, whereas it is undistillable whenever any other 
parties but $(A_3A_4)$ are joint. We choose $\rho_{III} \in \rho_4$ such that it 
is inseparable with respect to the bipartite splittings $(A_1A_2)$--$(A_3A_4)$, 
$(A_1)$--$(A_2A_3A_4)$ and $(A_2)$--$(A_1A_3A_4)$ and separable with respect to 
all other bipartite splittings.

The described activation effects can be understood using (ii) and (iii) of Sec. 
\ref{family}, together with the fact that when joining some of the parties, one 
may change the separability properties of certain bipartite splittings $P_k$ 
from separable to inseparable (see \cite{Du00} for details). 

We conclude that the experimental creation of non--maximally entangled 
$N$--partite states (not all $s_k=1$) might be of interest as well, as those 
states can have quite surprising properties. Note however that in this case, it 
is essential that the produced states are of the form $(\ref{rhoN})$, which can 
be accomplished by physically implementing the depolarization procedure 
described in \cite{Du99c}.


\subsection{Activating bound entanglement with bound entanglement}

Let us now consider the situation where the $N$ parties possess different kinds 
of BES, $\rho_1,\rho_2, \ldots \rho_L$, but this time remain spatially separated 
from each other. The parties again possess several copies of each of the states, 
i.e. $\rho_i^{\otimes M_i}$. By definition, it is clear that $\rho_i^{\otimes 
M_i}$ is not distillable for all $i$, i.e. the parties cannot create a MES if 
they have access to only one kind of BES. 

However if the parties have access to all different kinds of BES, i.e they share the state
\be
\rho'=\otimes_{i=1}^{L} \rho_i^{\otimes M_i},
\ee
we will give examples were they can distill a MES between some of the parties or 
even a GHZ state shared among all the parties. This effect, namely that the 
tensor product of two BES is no longer necessarily a BES was discovered by Shor 
{\it et al.} \cite{Sh00} and was termed ''Superactivation''. We shall refer to 
this as activation of bound entanglement with bound entanglement.

Let us investigate the simplest example of a tripartite system, $N=3$. We 
consider a state $\rho_1$ which is inseparable with respect to the bipartite 
splitting $A-BC$ and separable with respect to the splittings $B-AC$ and $C-AB$. 
As shown in \cite{Du00}, such a state is BE (a necessary condition for 
distillation of a MES shared between any two of the three parties is that at 
least two of the bipartite splittings have to be inseparable). Now consider 
states $\rho_2$ and $\rho_3$ which are created from the state $\rho_1$ by cyclic 
permutations of the parties, i.e. $\rho_2$ [$\rho_3$] is inseparable with 
respect to the splitting $B-AC$ [$C-AB$] respectively. For a particular choice 
of the states $\rho_1,\rho_2,\rho_3$, the parties can create -- once they have 
access to all three kinds of states -- a state $\tilde\rho$ which is inseparable 
with respect to all three bipartite splittings and which is in addition 
distillable to a GHZ state. In fact, they just have to pick randomly one of the 
three states $\rho_1,\rho_2,\rho_3$ (this can be accomplished via classical 
communication only), i.e.
\be
\tilde\rho=\frac{1}{3}(\rho_1+\rho_2+\rho_3).\label{aver}
\ee

To be specific, choosing $\rho_1$ within the family of states (\ref{rhoN}),$N=3$ with the following coefficients
\be
\lambda_0^+=\frac{1}{3}; ~~\lambda_0^-=\lambda_2=0; ~~\lambda_1=\lambda_3=\frac{1}{6}
\ee
ensures (i) that $\rho_1$ and $\rho_2,\rho_3$ (created by cyclic permutations of 
the parties) are BE with separability properties with respect to the 
bipartite splittings as announced above. (ii) $\tilde\rho$ defined in 
(\ref{aver}) is again of the form (\ref{rhoN}) with coefficients
\be
\tilde\lambda_0^+=\frac{1}{3}; ~~\tilde\lambda_0^-=0; ~~\tilde\lambda_1=\tilde\lambda_2=\tilde\lambda_3=\frac{1}{9},
\ee
and is inseparable with respect to all bipartite splittings (since $\Delta > 
2\lambda_k$) and hence distillable to a GHZ state.

It is now straightforward to extend these ideas to more parties and to a more 
general setup.  Therefore we consider a subfamily of $N$--qubit states of the 
form (\ref{rhoN}). We denote by $S$ all those bipartite splittings $P_k$ for 
which the state $\rho$ is inseparable (the corresponding $s_k=1$). For all other 
bipartite splittings $P_k \notin S$, $\rho$ is separable (the corresponding 
$s_k=0$). Let the number of separable bipartite splittings be $s>0$. We define 
$\Delta \equiv1/(s+1)$. The subfamily is defined by the following choice of 
parameters:
\bea
\label{subfamily}
\lambda_{0}^+=\Delta;&& ~~\lambda_{0}^-=0; \nonumber\\
\lambda_{k}=0 &\mbox{ iff }& P_k \in S \\
\lambda_{k}=\frac{\Delta}{2} &\mbox{ iff }& P_k \notin S.\nonumber
\eea
In general, we can announce the following

{\bf Theorem 1}: Given $L$ different kinds of BES $\rho_1, \ldots, \rho_L$ of the 
form (\ref{subfamily}), where $S_j$ denotes all bipartite splittings with respect 
to which $\rho_j$ is inseparable, one can create a state $\tilde\rho$ which is 
inseparable with respect to all those bipartite splittings where at least one of 
the states $\rho_j$ was inseparable, i.e. $\tilde S= \cup S_j$.
\\
{\it proof:} We define 
\be
\tilde\rho = \frac{1}{L} \sum_{j=1}^L \rho_j,
\ee
i.e. we pick randomly one of the states $\rho_j$ (which can be accomplished by 
classical communication) and show that $\tilde\rho$ has the desired properties. 
We have that $\tilde\rho$ is again of the form (\ref{rhoN}) and the coefficients 
$\tilde\lambda_k$ are given by the average of the coefficients $\lambda_{k,j}$ 
of the states $\rho_j$. We have to show that (i) $\tilde\Delta > 
2\tilde\lambda_k$ iff $P_k \in \tilde S\equiv \cup S_j$ and (ii) $\tilde\Delta 
\leq 2\tilde\lambda_k$ iff $P_k \notin \tilde S$. We have that 
$\tilde\Delta=1/L\sum_{j=1}^{L}\Delta_j$. In case (i), we have that least one of 
the states $\rho_j$ is inseparable with respect to the splitting $P_k$. We 
assume without loss of generality that it is only one, namely $\rho_1$ and thus 
$\lambda_{k,1}=0$ (the argument is exactly the same if more than one of the 
states $\rho_j$ are inseparable with respect to $P_k$). In this case we obtain 
for the corresponding $\tilde\lambda_k=1/L\sum_{j=2}^{L}\Delta_j/2$. Note that 
the sum runs from $j=2$ to $L$, which ensures that (i) is fulfilled ---since 
$(\tilde\Delta-2\tilde\lambda_k)=\Delta_1/L>0$. In case of (ii), i.e. $P_k 
\notin S_j~\forall j$, we find $\tilde\lambda=1/L\sum_{j=1}^{L}\Delta_j/2$ and 
(ii) is fulfilled, which finishes the proof of our statement.

Given this theorem, it is now very easy to construct several examples which show the 
activation of bound entanglement with bound entanglement:

{\bf Example 1:} 
We consider $N$ parties and assume that $N$ is even. We have $N/2$ different BE 
states $\{\rho_k\}, k=1,2,\ldots,N/2$. If the parties have access to any 
$(N/2-1)$ (or less) different kinds of BE states $\rho_k$, they cannot distill 
any entanglement. However, once the parties have access to all kinds of BE 
states $\rho_k$, they can create a state $\tilde\rho$ which is inseparable with 
respect to all bipartite splittings and thus distillable to a $N$--party GHZ 
state.  The following choice of states has the announced properties: The state 
$\rho_k$ is of the form (\ref{subfamily}) and is inseparable with respect to all 
bipartite splittings which contain exactly $k$ parties on one side and $N-k$ 
parties on the other side and separable with respect to all other bipartite 
splittings. This ensures that all state $\rho_k$ are BE \cite{Du00} and 
---according to Theorem 1--- the parties can create a state $\tilde\rho$ which 
is distillable to a GHZ state once they have access to all $N/2$ different 
states $\rho_k$. If the access is limited to $(N/2-1)$ or less different kinds 
of BE states $\rho_k$, one can easily check using (iii) of Sec. \ref{family} 
that no entanglement can be distilled.

{\bf Example 2:} 
In this example, we consider $N$ different BE states $\{\rho_l\}, 
l=1,2,\ldots,N$. Here, the state $\rho_N$ serves as a ''key-state'', as on one 
hand, access to $\rho_N$ together with access to the state $\rho_l$ enables the 
parties $A_l$ and $A_N$ to distill a MES. On the other hand, access to all 
states $\rho_l$ except $\rho_N$ does not allow the parties to distill any 
entanglement at all. If in addition also $\rho_N$ is accessible, a GHZ state 
shared by all the parties can be distilled (as party $A_N$ can create a MES 
shared with any party $A_l$). Such a situation can be established by the 
following choice of states: For $l \not= N$, the state $\rho_l$ is of the form 
(\ref{subfamily}) and is inseparable with respect to all bipartite splittings 
which have parties $A_l$ and $A_N$ on different sides, except the splittings 
$A_l$--rest and $A_N$--rest which as well as all the other splittings are 
separable. The state $\rho_N$ is also of the form (\ref{subfamily}) and is 
inseparable with respect to all splittings where exactly one particle is on one 
side and $N-1$ particles are on the other side. All states $\rho_l$ are BE, 
which can be checked using (ii) and (iii) of Sec. \ref{family}. 
Applying Theorem 1, it is easy to observe the described activation effect.

Note that the activation of BE by joint actions may be combined with the 
activation of BE with BE. This opens a huge variety of different examples, which 
can all be constructed using the results of Ref. 
\cite{Du00} together with Theorem 1 and states of the form (\ref{subfamily}).      

\section{Summary}\label{Summary}

We discussed several aspects of multipartite entanglement and its experimental 
detection. First we focused on bipartite aspects of MPE, which can be determined 
by investigating the bipartite reduced density operators of the multipartite 
systems. We then used $l$--partite splittings to establish the $l$--separability 
and distillability properties of a multipartite density operator $\rho$. For a 
certain family of states, we completely determined the separability and 
distillability properties using bipartite splittings only. Using this, we 
provided a simple method to determine whether a mixed state $\rho$ is 
multipartite entangled, and in addition to detect which kind of entanglement is 
present. We illustrated this method by revisiting two recent experiments. 
Finally, we focused on bound entangled states and the activation of BE. We 
showed that BE can be activated by joint actions of the parties or with help of 
a different kind of BE itself.

\section{Acknowledgements}

We thank G. Vidal and J. Smolin for discussions. This work was supported by the Austrian 
Science Foundation under the SFB ``control and measurement of coherent quantum 
systems´´ (Project 11), the European Community under the TMR network 
ERB--FMRX--CT96--0087 and project EQUIP (contract IST-1999-11053), the European 
Science Foundation and the Institute for Quantum Information GmbH.




\begin{references}
\bibitem{noteMol}
Note that the inseparability of a certain reduced density operator $\rho_{kl}$ 
already implies that a MES ---such as 
$|\Phi^+\rangle=1/\sqrt{2}(|00\rangle+|11\rangle)$--- shared between parties 
$A_k$ and $A_l$ can be distilled (when allowing for several copies of the 
state), even without help of the remaining parties. This is due to the fact that 
for two qubit systems, inseparability is equivalent to distillability 
\cite{Ho97}. In fact, the remaining parties can by no means prevent parties 
$A_k$ and $A_l$ to distill a MES.
\bibitem{noteExp}
In \cite{Sa00}, only the probabilities $p_l$ to find $l$ ions in state 
$|0\rangle$ and $N-l$ ions in state $|1\rangle$ are given. For $l=1$, there are 
4 states which might contribute to $p_1$, namely $|\phi^1_i\rangle \in 
\{|0111\rangle,|1011\rangle,|1101\rangle,|1110\rangle\}$. From the measured date 
we know that $p_1=P_1\rho P_1=$ tr$(\sum 
d^1_{ij}|\phi^1_i\rangle\langle\phi^1_j|)$, where $P_1=\sum 
|\phi^1_j\rangle\langle\phi^1_j|$ and we assumed that $d^1_{ij}$ are the 
(unknown) coefficients of the density matrix $\rho$ in the basis 
$|\phi^1_i\rangle$.  Similarly, for $l=3$, the four states $|\phi^3_i\rangle \in 
\{|0001\rangle,|0010\rangle,|0100\rangle,|1000\rangle\}$ can contribute to $p_3$ 
and tr$(\sum d^3_{ij}|\phi^3_i\rangle\langle\phi^3_j|)=p_3$. For $l=2$, there 
are 6 states which might contribute to $p_2$, namely $|\phi^2_i\rangle \in 
\{|0011\rangle,|0101\rangle,|0110\rangle,|1001\rangle,|1010\rangle, 
|1100\rangle\}$ and again tr$(\sum 
d^3_{ij}|\phi^3_i\rangle\langle\phi^3_j|)=p_3$. Note that the coefficients 
$d_{ij}^l$ are not specified by the provided data, but the diagonal coefficients 
$d^l_{ii}$ are bound from above by $p_l$. We have e.g. that 
$2\lambda_{001}=d^1_{11}+d^3_{22}$. Since we might get contributions to 
$\lambda_{001}$ from $p_1$ and $p_3$, we conclude that $0\leq 2\lambda_{001} 
\leq p_1+p_3$. A similar argument holds e.g. for 
$2\lambda_{110}=d^2_{66}+d^2_{11}$, but in this case we only get contributions 
from $p_2$. We thus conclude that $0\leq 2\lambda_{110}\leq p_2$. The same line 
of argumentation can be used to obtain bounds for all $\lambda_k$. In the 
experiment, the values of $p_j$ were determined to be $p_1=p_2=p_3=0.10 \pm 0.02$.
\end{references}
\end{document}